\documentclass[aps, prapplied, twocolumn,10pt,showpacs,superscriptaddress,amsmath,amssymb,nobibnotes]{revtex4-1}
\usepackage{graphicx}
\usepackage{amsfonts}    
\usepackage{verbatim}   
\usepackage{color}      
\usepackage{subfigure}  
\usepackage{hyperref}   
\usepackage{booktabs}
\usepackage{threeparttable}
\usepackage{dcolumn}
\usepackage{multirow}
\usepackage{physics}
\usepackage{colortbl}
\definecolor{tabcolor}{rgb}{0,0,0}
\usepackage{upgreek}
\begin{document}

\title{Induced-photorefractive attack against Quantum Key Distribution}

\author{Peng Ye}
\author{Wei Chen}\email{weich@ustc.edu.cn}
\author{Guo-Wei Zhang}
\author{Feng-Yu Lu}
\author{Fang-Xiang Wang}
\author{Guan-Zhong Huang}

\author{Shuang Wang}
\author{De-Yong He}
\author{Zhen-Qiang Yin}

\author{Guang-Can Guo}
\author{Zheng-Fu Han}

\affiliation{CAS Key Laboratory of Quantum Information, University of Science and Technology of China, Hefei 230026, China}
\affiliation{CAS Center for Excellence in Quantum Information and Quantum Physics, University of Science and Technology of China, Hefei 230026, China}

\date{\today}

\begin{abstract}
\noindent 
Lithium niobate ($\rm LiNbO_{3}$, LN) devices play critical roles in quantum information processing. However, for special applications like quantum key distribution (QKD), the characteristics of materials and devices and their impact on practical systems must be intensively inquired. For the first time, we reveal that the photorefractive effect in LN can be utilized as a potential loophole to carry out malicious attacks by the eavesdroppers. We take a commercial LN-based variable optical attenuator as an example to demonstrate the method we named Induced-photorefractive attack (IPA) and propose two techniques to enable controllable attacks. Our results show that eavesdroppers can fulfill an efficient source-side attack by injecting an optimized irradiation beam with only several nanowatts, which is realistic when accessing commercial fiber channels. These measure and techniques can be employed for all individual and on-chip LN devices and initially explored a new security branch for system design and standardization of real-life QKD.

\end{abstract}


\maketitle

\section{Introduction}
The photoinduced index change procedure in lithium niobate (LN) is known as the photorefractive effect (PE), which causes a phase shift between the illuminating profile and the index pattern \cite{1985hallPiQE}. As an inherent property of LN \cite{1978glassOE}, PE can remarkably change the index of LN waveguide, which further influences a series of modulation results such as phase, intensity, polarization, or extinction ratio \cite{mondain2020photorefractive,harvey1988photorefractive} according to different device structures and make them no longer reliable. Although many studies have been conducted on the imperfections of LN devices \cite{2018yoshinonQI,2019weiPRA}, research on the security problems associated with the index change of material has not been in-depth concerned from the perspective of practical security in QKD systems. 

For the first time, we propose that the PE can be utilized by eavesdroppers (Eve) to actively modify the parameters of QKD systems and manipulate the results of quantum state preparation, which may result in a totally broken security at worst. We analyze the mechanism of PE and the effect of the induced-photorefractive attack (IPA). We remark that the PE is a new mechanism to cause loopholes and can be applied to affect almost all LN devices, such as intensity modulators (IM), phase modulators (PM), polarization controllers (PC), variable optical attenuators (VOA), and choppers since PE is an inherent property of LN material. The implementation methods of attacks may differ for different types of LN devices. Here, a LN-based VOA in QKD transmitters is used as an example to verify IPA and two techniques are proposed to enable precisely controlled attacks. Simulation results show that Eve can completely steal the secure key without being perceived if she can control the attenuation of VOA. The experiment demonstrates that Eve can commit private hacking with just 3 nanowatts irradiation power, which is 9 orders of magnitude lower than the laser-damage scheme \cite{huang2020laser}. This indicates IPA can be executed from the available accessing points in real-life fiber channels. 

In addition to fiber devices, the PE is also present and more evident in chip-based LN devices owing to the stronger confinement of optical mode in lithium niobate on insulator (LNOI) waveguides \cite{2018boesLPR, 2020linPRP,2021zhuAOPA}. As we know. the LNOI-based high-speed modulators not only have a higher modulation bandwidth \cite{2022xueOO,wang2018integrated} but also do not have the modulation-dependent loss \cite{PhysRevA.105.012421,2022yeOE} thanks to the Pockels effect \cite{2015liuLPR}. It matches the requirements of QKD very well from both the modulation and security. We believe that LNOI has great potential and application prospects in integrated QKD in the future thus the PE-related issues shouldn't be ignored either.
According to our results, the presence of PE should be seriously considered when designing QKD systems, and this work can give essential inspiration to enhance their practical security.


\section{The mechanism of IPA} \label{sec:mechanism}
\begin{figure}[htbp]
	\includegraphics[width=8.5cm]{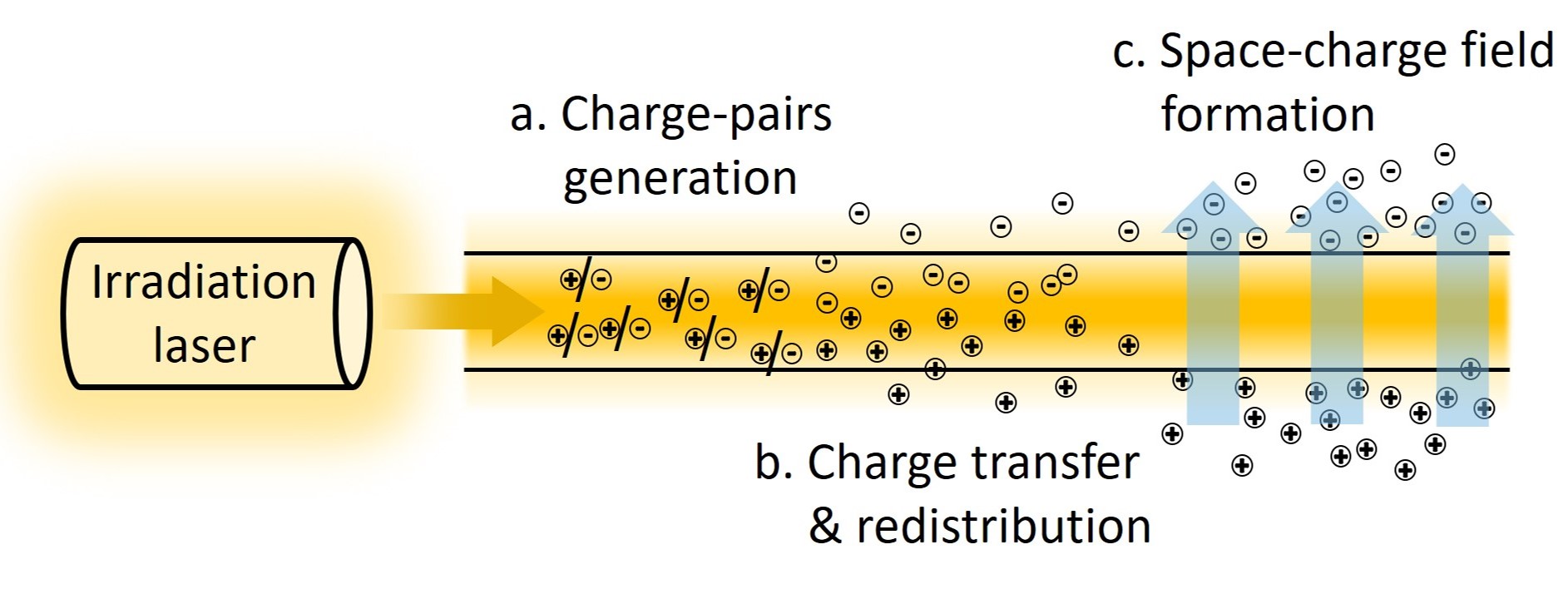}
	\caption{\label{fig: PEprinciple} Schematic diagram of PE's procedure.}
\end{figure}

As shown in Fig. \ref{fig: PEprinciple}, the PE procedure can be described in three stages: (a) Charge-pairs generation: The irradiation beam excites photogenerated charge pairs from impurities and defect centers. (b) Charge transfer and redistribution: The charges are driven by the carrier concentration distribution, the electric field, or the photovoltaic effect thus they move out of the waveguide and get trapped in dark areas. (c) Space-charge field formation: The inhomogeneous charge distribution causes a space-charge field, which modulates the index of LN waveguides through the Pockels effect \cite{kostritskii2009photorefractive}.

Under irradiation power $I_{\mathrm{ir}}$, the photoinduced index change can be derived from the Kukhtarev equation \cite{1978kukhtarevF}:
\begin{equation}
\Delta n(I_{\mathrm{ir}},t) = \frac{n^{3}r_{\mathrm{33}}\gamma }{2}E_{\mathrm{s}}(I_{\mathrm{ir}},t )=\Delta n_{\mathrm{s}}(I_{\mathrm{ir} })(1-e^{-\frac{t}{\tau} }),
\label{eq:nt}
\end{equation}
where $E_{\mathrm{s}}(I_{\mathrm{ir}},t )$ is the space-charge field induced by PE, $n$ is the waveguide refractive index at signal wavelength, $r_{33}$ is the Pockels coefficient, $\gamma $ is the overlap integral of the optical mode profile and $E_{\mathrm{s}}$. $\tau$ is the build-up time constant \cite{fujiwara1993comparison}, which determines the response speed of PE and can be expressed by $\tau=\epsilon _{\mathrm{r}}\epsilon _{0}/\left (\sigma _{\mathrm{d}}+\sigma _{\mathrm{ph}}  \right )$, where $\epsilon _{\mathrm{r}}$ and $\epsilon _{0}$ are relative and vacuum dielectric constant respectively, $\sigma _{\mathrm{d}}$ is the dark conductivity and $\sigma _{\mathrm{ph}}$ is the photoconductivity positively correlated with $I_{\mathrm{ir}}$. The saturated photorefractive index change $\Delta n_{\mathrm{s}}(I_{\mathrm{ir}})$ is \cite{becker1985photorefractive} 
\begin{equation}
\Delta n_{\mathrm{s}}(I_{\mathrm{ir}})=\frac{n^{3}r_{33} \gamma}{2 }(-\frac{\sigma _{\mathrm{ph}}}{\sigma}E_{\mathrm{app}}+\frac{\kappa \alpha}{\sigma} I_{\mathrm{ir}}),
\label{eq:thetaPEt}
\end{equation}
where $E_{\mathrm{app}}$ is the applied electric field, $\alpha$ is the absorption coefficient of the donor or acceptor centers, $\sigma =\sigma _{\mathrm{d}}+\sigma _{\mathrm{ph}} $, and $\kappa$ is the Glass constant. See Appendix \ref{App:PEdetail} for more details about PE.

As we mentioned before, this index change causes the signal phase to change by $\theta _{\mathrm{PE}}=2\pi  \Delta n_{\mathrm{s}}L_{\mathrm{eff}} /\lambda _{0} $, where $L_{\mathrm{eff}}$ is the effective interaction length and $\lambda _{0}$ is the wavelength of signal photons, and then a series of modulation deviations in LN devices for signal photons. For example, the signal quantum state passing through a Mach-Zehnder interferometer (MZI) is
\begin{equation}
|\sqrt{\mu}\rangle=|i\sqrt{r (1-r )}(e^{i\Delta \theta}+1)\sqrt{\mu _{\mathrm{in}}}\rangle,
\label{eq:Iout}
\end{equation}
where $\mu _{\mathrm{in}}$ is the mean photon number (MPN) of the input signal pulses, $r$ is the beam-splitting ratio (BSR) of two beam splitters (BS) in MZI. The phase difference between the two arms is expressed as $\Delta\theta = \theta _{\mathrm{arm1}}-\theta _{\mathrm{arm2}}$. We can see that the index change of the LN waveguide under IPA influences both $r$ and $\Delta\theta$ and then the signal quantum state $|\sqrt{\mu}\rangle$. 

Some researchers consider that the PE in MZI only comes from the change of $r$ because the PE on two arms can cancel each other \cite{harvey1988photorefractive}. 
But in fact, since the fabrication error and the applied modulation electric fields $E_{\mathrm{app}}^{\mathrm{i}}$ ($i=1$ for the upper arm and $i=2$ for the lower arm) are both different for the two arms of MZI, their working conditions are not identical. Additionally, the $I_{\mathrm{ir}}^{\mathrm{i}}$ in these two arms are also different because the $r$ deviated from 0.5 at the irradiation wavelength. According to Eq. \ref{eq:thetaPEt}, the differences in both $E_{\mathrm{app}}^{\mathrm{i}}$ and $I_{\mathrm{ir}}^{\mathrm{i}}$ on each arm leads to the inconsistent PE so that they cannot cancel each other. In contrast, the effect of $r$ caused by PE is almost negligible at low irradiation power ($I_{\mathrm{ir}} <$100 mW) \cite{kostritskii2009photorefractive}. Hence we deem that $\Delta\theta$ dominates the attenuation change.

The phase difference $\Delta\theta$ can be divided into three parts \cite{howerton1992photorefractive,becker1985photorefractive}, which can be described as
\begin{equation}
\Delta\theta = \Delta\theta _{0}+\Delta\theta_{\mathrm{E}}+\Delta\theta_{\mathrm{PE}},
\label{eq:thetaTotal}
\end{equation}
including a geometrical phase $\Delta\theta _{0}$, an electro-optic phase $\Delta\theta_{\mathrm{E}}$ and a PE phase $\Delta\theta_{\mathrm{PE}}$, which corresponds to the structural difference coming from the fabrication error due to technological imperfections on each arm, the difference of modulation voltages applied to each arm, and the difference of photoinduced index change between two arms, respectively. Among them, the $\Delta\theta_{\mathrm{E}}$ can be expressed as
\begin{equation}
\Delta\theta_{\mathrm{E}}=\pi \frac{\Delta V}{V_{\pi }},
\label{eq:thetaE}
\end{equation}
where $V_{\pi }$ is the driving voltage required for the phase modulator to provide $\pi$ phase. $\Delta V=V_{\mathrm{arm1}}-V_{\mathrm{arm2}}$  refers to the modulated voltage difference between two arms. We assume $V_{\mathrm{arm1}}=-V_{\mathrm{arm2}}=V_{\mathrm{app}}$  for electrode working at push-pull mode.

The PE phase $\Delta\theta_{\mathrm{PE}}$ is equal to $ \theta_{\mathrm{PE}}^1(I_{\mathrm{ir}}^{1} )- \theta_{\mathrm{PE}}^2 (I_{\mathrm{ir}}^{2} )$. Then we can calculate the total phase difference between two arms as
\begin{equation}
\begin{split}
\Delta \theta=&\Delta \theta _{0}+V_{\mathrm{app}}\left [ \frac{2\pi }{V_{\pi }}-\frac{C}{d}\Delta^{+}f(I_{\mathrm{ir}})\right ]+D \Delta ^{-}f(I_{\mathrm{ir}}).
\end{split}
\label{eq:thetaPE}
\end{equation}
This equation applies to any MZI-based LN devices and has been simplified by introducing $C = 2  a \pi   L_{\mathrm{E}}/\kappa \lambda _{0}$ and $D = 2 \pi  L/\lambda _{0}$, where $a$ is a constant \cite{fujiwara1993comparison}, $\lambda_{0}$ is the free-space wavelength of signal laser, $L$ is the length of arm. We also define a function $ f(I_{\mathrm{ir}}^{i})=AI_{\mathrm{ir}}^{i}/(1+BI_{\mathrm{ir}}^{i})$, which is the index changed only depending on $I_{\mathrm{ir}}$ \cite{fujiwara1993comparison}, where $A = n^{3}r_{33} \gamma\kappa a/2\sigma _{\mathrm{d}}$ and $B = a\alpha/\sigma _{\mathrm{d}}$. Due to the different PE on each arm, we have \begin{equation}
\begin{split}
\left\{\begin{matrix}
\Delta ^{+}f(I_{\mathrm{ir}})=f(I_{\mathrm{ir}}^{1})+f(I_{\mathrm{ir}}^{2})\vspace{1ex}\\ 
\Delta ^{-}f(I_{\mathrm{ir}})=f(I_{\mathrm{ir}}^{1})-f(I_{\mathrm{ir}}^{2})
\end{matrix}\right..
\end{split}
\end{equation}

Therefore, when we use a MZI-based VOA to analyze the effect of IPA, the $\mu$ in Eq. \ref{eq:thetaPE} is the MPN of signal pulses after attenuation and we can see that it is affected by both $I_{\mathrm{ir}}$ and $V_{\mathrm{app}}$. This enables Eve to control the attenuation and execute IPA.

\section{Experiment}
\begin{figure}[htbp]
	\includegraphics[width=8.5cm]{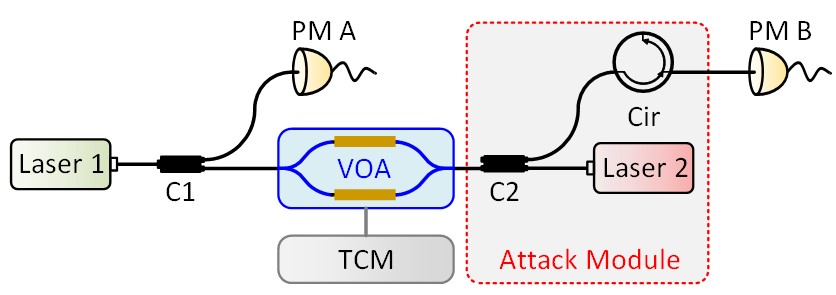}
	\caption{\label{fig:ExperimentSetup}Sketch of the experimental setups. Laser 1: a 1550 nm signal laser (Agilent 81960A). C1: coupler 1, a fiber 1:99 polarization-maintaining BS. PM A: power meter A (Dwin BD613C). VOA: the schematic diagram of the internal structure of the variable optical attenuator (iXblue MXAN-LN-10). The blue lines and yellow blocks here are LN waveguides and modulating electrodes, respectively. TCM: a temperature control module (Oeshine TCM-M207). C2: coupler 2, a fiber 50:50 BS. Cir: a fiber circulator. Laser 2: the 405 nm irradiation laser (Max-Ray FL-405-20-SM-B). PM B: (Thorlabs PM100D with S132C).}
\end{figure}

As shown in Fig. \ref{fig:ExperimentSetup}, we design experiments to verify the effect of IPA on VOA.
Laser 1 is a 1550 nm signal laser monitored by C1 and PM A. The device we test is a commercial MZI-based VOA, whose temperature is controlled at 30 $^{\circ}$C with the accuracy of 0.01 $^{\circ}$C by the TCM. Since this LN device is polarization-dependent, the polarization-maintaining fibers are used here to connect these components. The attack module (see Appendix \ref{App:AttModule} for more details) is used to insert the channel and execute the IPA. In this module, Laser 2 is the irradiation laser working at 405 nm due to higher photorefractive sensitivity at a shorter wavelength \cite{fujiwara1989wavelength} and its power is tunable from 0 to 20 mW. Here we achieve the inverse injection of the irradiation beam into VOA through C2 and the Cir is used to block the reflected irradiation beam from entering PM B, which monitors the effect of IPA.

\begin{figure*}[htbp]
	\includegraphics[width=18cm]{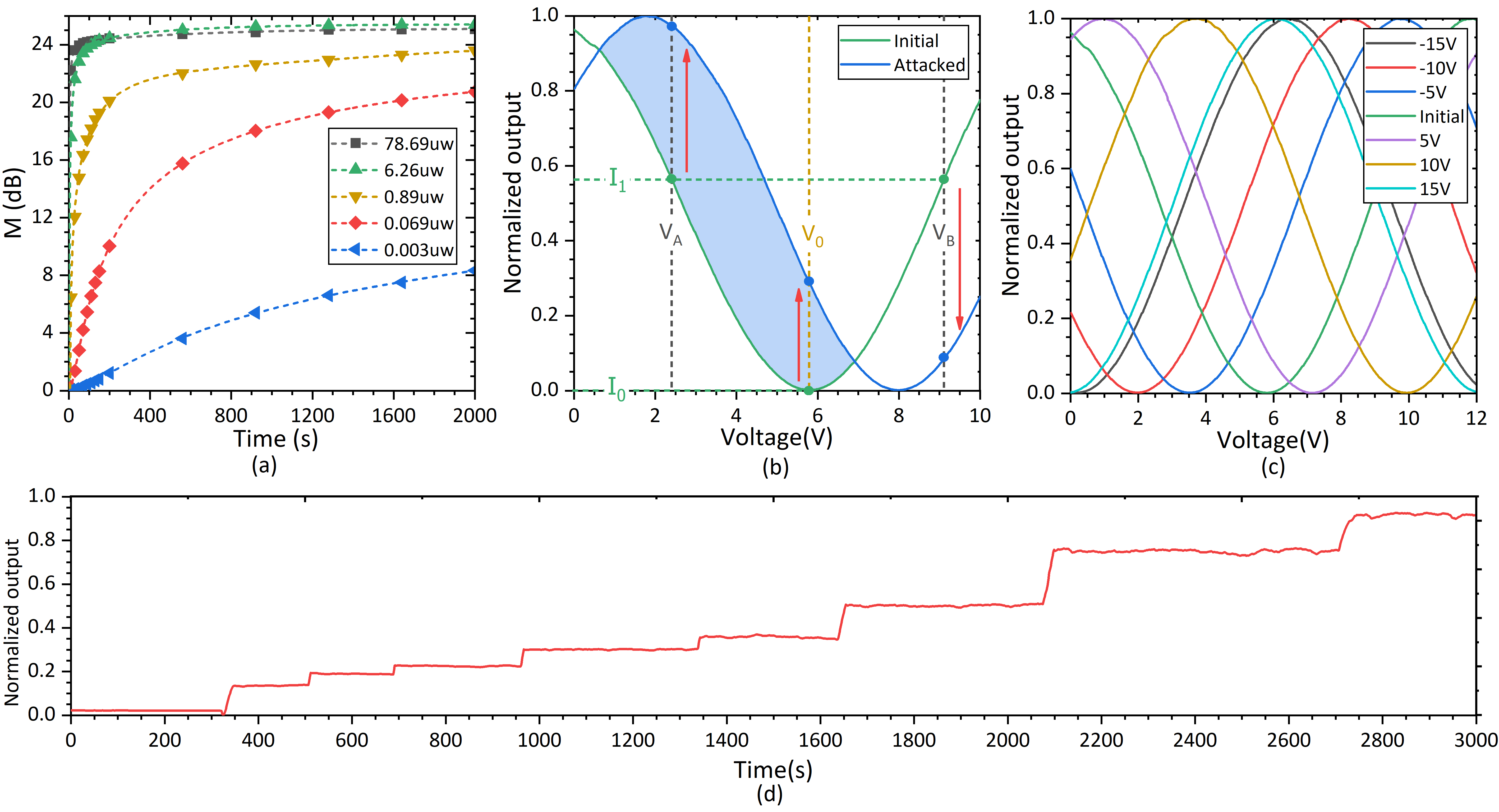}
	\caption{\label{fig:AllExpResult} (a) The variation of $M$ obtained by dividing the attacked signal intensity by the original one is shown here in logarithmic coordinates on the Y-axis. (b) The schematic of IPA at different working voltages. (c) The result of the pre-treatment technique with 12 $\upmu$W irradiation power under different applied electric field. (d) The result of the pulse-injection technique with 12 $\upmu$W peak power and 10-second period pulses.}
\end{figure*}

To analyze how IPA impacts QKD, the signal intensity change is represented by magnification $M$ obtained by dividing the attacked signal intensity by the original ones and we test it for different irradiation powers (Fig. \ref{fig:AllExpResult}a). Here we set the original state of VOA at its maximum attenuation (i.e., its minimum output $I_0$ at working voltage $V_0 = 5.8$ V) to demonstrate IPA more clearly. Under IPA, VOA’s attenuation decreases and this process corresponds to the change from the green point to the blue one along the dotted line of $V_0$ in Fig. \ref{fig:AllExpResult}b. 
As shown in Fig. \ref{fig:AllExpResult}a, the index change gradually saturates as the injection duration and then we get the saturated $M$. As we analyzed, both the saturated $M$ and the response speed increase at higher irradiation powers, which corresponds to the increase of $\Delta n_{\mathrm{s}}(I_{\mathrm{ir}} )$ and $1/\tau$ respectively.

In Fig. \ref{fig:AllExpResult}a, the saturated $M$ decreases slightly when the irradiation power is larger than 6.26 $\upmu$W. This essentially stems from the saturation property of PE, i.e., $\Delta n_{s}(I_{\mathrm{ir}} )$ also tends to saturation \cite{fujiwara1993comparison}.
Since $I_{\mathrm{ir}}^{1}\neq I_{\mathrm{ir}}^{2}$, with the increase of $I_{\mathrm{ir}}$, the index change of the arm with higher irradiation power reaches saturation first while that in another arm still increases. This results in a decrease in phase difference between two arms and consequently the decrease in the saturated $M$. 
We can see that VOA's attenuation can be impacted even at only 3 nW irradiation power. We note that these powers are measured at the connection of VOA and C2. Therefore, the irradiation power actually entering the LN waveguide is lower than what we measured considering the couple loss between the fiber and the LN chip inside the VOA package. 

Considering more general cases, VOA's output may not be the minimum value while some intermediate value (such as $I_1$ in Fig. \ref{fig:AllExpResult}b) is more common. In this case, there are two working states of VOA that can output $I_1$, corresponding to working on the falling ($V_{\mathrm{A}}$) and rising ($V_{\mathrm{B}}$) edges of its voltage curve. Under IPA, the voltage curve of VOA shifts from the initial (green) line to the attacked (blue) one. Our attack is working (e.g. $V_{\mathrm{A}}$) as long as the attacked voltage curve is above the initial one (the blue-shaded area). However, the working state of VOA is sometimes not suitable for attacks. For example, the attenuation increases instead when the working voltage (e.g. $V_{\mathrm{B}}$) is outside the blue-shaded area.


To make IPA available, the user-calibrated working voltage should be located in the blue area. In addition, since PE is voltage-dependent (Eq. \ref{eq:thetaPEt}), it is better to have a working voltage as high as possible to produce larger PE.  These objectives can be achieved by the pre-treatment technique. According to Eq. \ref{eq:thetaPEt}, we treat LN devices by injecting the irradiation beam until saturation under appropriate $E_{\mathrm{app}}$ and $I_{\mathrm{ir}}$ and then turn off the applied electric field and irradiation beam simultaneously. Considering the PE in LN without the applied electric field or any other treatments always persists for several weeks or longer \cite{1978glassOE}. The pre-treatment results can remain in LN devices long enough to affect their working states in QKD systems. As shown in Fig. \ref{fig:AllExpResult}c, by using the irradiation laser at continuous-wave mode and applying $I_{\mathrm{ir}}$ at 12 $\upmu$W with different $V_{\mathrm{app}}$, we can shift the static bias of VOA's voltage curve to any position of the entire $2\pi$ range. Based on this technology, Eve can preset the original state of VOA to the most beneficial state for IPA before delivering it to users.

Furthermore, we propose the pulse-injection technique to enable fast, accurate, and arbitrary controllable attacks. This technique is performed by operating the irradiation laser at the pulse mode. In this technique, fast response (small $\tau$) and significant index change (large $\Delta n_{\mathrm{s}}(I_{\mathrm{ir}})$) can be achieved by utilizing sufficiently high peak power pulses, while arbitrary precision manipulation can be ensured by controlling the duty cycle and frequency of injected pulses. 
In our experiment, the peak power of pulses is 12 $\upmu$W. To demonstrate this technique more apparent, we use 10-second period pulses and deliberately perform multiple injections. 
During the injection, we use the 1-second width pulses to change the output of VOA. When the desired extent of IPA is reached, we decrease the duty cycle of pulses to stabilize the effect of our attack, which is essentially to offset the decay behavior of PE \cite{fujiwara1989evolution}.

As shown in Fig. \ref{fig:AllExpResult}d, by finely tuning the duty cycle of pulses, the extent of IPA can be controlled at will and the effect of IPA can be well stabilized. To demonstrate the effect of pulse injection technology more clearly, the maximum attenuation state of VOA is processed to around 0V through the pre-treatment technique at -15 V applied electric field with 12 $\upmu$W irradiation power. According to Fig. \ref{fig:AllExpResult}c, the effect of IPA would be further enhanced if an applied electric field is present during this test. Therefore, to demonstrate the effect of the irradiation beam itself only, here we did not apply any external electric field hence the output of VOA cannot be modulated to its maximum attenuation state exactly and this is why the starting point of Fig. 3d in main text is slightly deviated from the minimum value. In other words, under the current conditions, here we show only the lower limit of the effect for IPA.
In practical attack scenarios, faster and more accurate attacks can be achieved by programmed automatic feedback control of the irradiation laser. More details about experiments are shown in Appendix \ref{App:Expmethod}.

\section{The IPA against BB84 QKD}
We use the decoy state BB84 protocol \cite{2005maPRA} to analyze the security suffering from IPA (i.e., the $M$).
In our simulation, the overall transmittance of Bob's detection apparatus is 0.1, the fiber loss is 0.2 dB/km, the probability that a photon hit the erroneous detector is 0.5\%, and the dark count rate is $6\times 10^{-7}$. At the transmitter, Alice sends weak laser pulses with Poissonian photon number distribution and the MPN of these signal pulses and decoy pulses are attenuated to $\mu$ and $\nu$, respectively. The total transmittance between Alice and Bob is $\eta$, which can be regarded as the product of the transmission of the quantum channel $\eta_{\mathrm{AB}}=10^{-\alpha L/10}$ and $\eta_{\mathrm{B}}$ (the internal efficiency of Bob). 
When there is no eavesdropper, the MPN of signal pulses at Bob is $\eta\mu$ and still Poissonian.
The response rate of signal state at Bob is $Q_{\mu }=Y_{0}+1-e^{-\eta \mu }$,
where $Y_{0}$ is the dark count rate. Under the IPA, the attenuation of VOA is degraded so that the attacked MPN of the signal state is $\mu_{\mathrm{e}} = M\cdot \mu$, namely, the MPN is amplified by $M$ times for both signal and decoy states. 
Then Eve intercepts all pulses from Alice and resents each photon to Bob with probability $p$. The remaining photons not sent to Bob are stored in quantum memory. As a reminder, there is no need for Eve to have the ability to distinguish the photon number in each pulse. The probability of a successful attack can be defined by that Eve stores and resents at least one photon which can be detected by Bob, that is 
\begin{equation}
\begin{split}
p_\mathrm{s} = \sum_{n=2 }^{\infty}P_{\mu_{\mathrm{e}}}(n)\sum_{m=1}^{n-1}C_{n}^{m}p^{m}(1-p)^{n-m}\left [ 1-\left ( 1-\eta _{\mathrm{B}} \right )^{m} \right ]
\end{split}
\end{equation}

Eve's attack is always defined as a failure for pulses with only one photon, even if it has a probability of $1-p$ to be intercepted. This operation causes the photon number distribution change to 
\begin{equation}
\begin{split}
P_{\mathrm{PNS}}(n)=&\sum_{ m=n}^{\infty}P_{\mu_{\mathrm{E}}}(m)(1-p)^{m}(\frac{p}{1-p})^{n}C_{m}^{n}\\
=& e^{-Mp\mu }\frac{(Mp\mu )^{n}}{n!},
\end{split}
\end{equation}
which is still Poissonian distribution and the MPN is $Mp\mu$. At the internal efficiency $\eta_\mathrm{B}$ at Bob, we can get the response rate of pulses with $n$ photons is $Y^{\mathrm{PNS}}_n=Y_{0}+1-(1-\eta_{\mathrm{B}} )^{n}(n\geq 1)$. Then the response rate of the signal state after being attacked can be detected as
\begin{equation}
\begin{split}
Q_{\mu }^{'}=\sum_{\mathrm{n=1}}^{\infty }P_{\mathrm{PNS}}(\mathrm{n})Y^{\mathrm{PNS}}_\mathrm{n} =Y_{0}+1-e^{-M \mu p \eta_{\mathrm{B}}  }.
\end{split}
\end{equation}

To avoid disturbing the count rate of signal states detected by Bob, Eve should keep $Q_{\mu }^{'}=Q_{\mu }$, that is
\begin{equation}
p =\frac{\eta_{\mathrm{AB}}}{M}.
\end{equation}

For decoy-state pulses, we can get the same results for $Q_{\nu }^{'}=Q_{\nu }$. In other words, if Eve intercepts all pulses and resents each photon to Bob by probability $p =\eta_{\mathrm{AB}}/M$, she will not be detected through monitoring the count rate of signal and decoy states. After the public discussion part of the QKD protocol, Eve can get the same result as Bob by measuring the stored signal photons on the same basis. 

Due to the undisturbed response rate on signal and decoy states, the tagged bits rate estimated by legal users is \cite{GLLP}
\begin{equation}
\Delta=\frac{Q_{\mu }-P_{0}Y_{0}-P_{1}Y_{1}}{Q_{\mu }}\approx 1-\frac{P_{1}Y_{1}}{Q_{\mu }},
\end{equation}
where $Y_1$ is the lower bound of the single-photon yield, $Q_{\mu }$ is the response rate of signal state pulses with MPN $\mu$, and $P_{\mathrm{i}}$ of $i\in \left \{ 0,1 \right \}$ is the i-photon probability of the Poissonian distribution pulses. However, the actual $\Delta$ is affected by IPA, that is, 
\begin{equation}
\begin{split}
\Delta^{\mathrm{PNS}} = \sum_{\mathrm{n}=2 }^{\infty}\frac{P_{\mu_{\mathrm{e}}}(\mathrm{n})}{Q_{\mu }} \sum_{\mathrm{m}=1}^{\mathrm{n}-1}C_{\mathrm{n}}^{\mathrm{m}}p^{\mathrm{m}}(1-p)^{\mathrm{n}-\mathrm{m}}\left [ 1-\left ( 1-\eta _{\mathrm{B}} \right )^{\mathrm{m}} \right ]
\end{split},
\end{equation}
where $\mu_{\mathrm{e}}$ is the MPN of signal states suffering from the IPA and expressed as $\mu_{\mathrm{e}} = M\cdot \mu$, 
$\eta_{\mathrm{B}}$ is the internal efficiency of Bob. The secret key per pulse is calculated by \cite{2005wangPRL}
\begin{equation}
R = Q_{\mu }\left \{(1-\Delta^{\mathrm{PNS}} )[1-H_{2}(e_{1})]-f H_{2}(E_{\mu })  \right \}
\end{equation}
where $H(x) = -x\log_{2}x-(1-x)\log_{2}(1-x)$ is the Shannon's binary entropy function, $e_1$ is the upper bound of the single-photon quantum bit error rate (QBER) estimated by the decoy-state method, $f = 1.16$ is the error correction efficiency, $E_{\mu }$ is the overall signal states QBER.
The attack does not change the counting rate of the signal and decoy states, and both $e_1$ and $E_\mu$ are identical as no attack, which means that Eve can partially or even completely steal the security key without being discovered by users. 

\begin{figure}[htbp]
	\centering\includegraphics[width=0.9\linewidth]{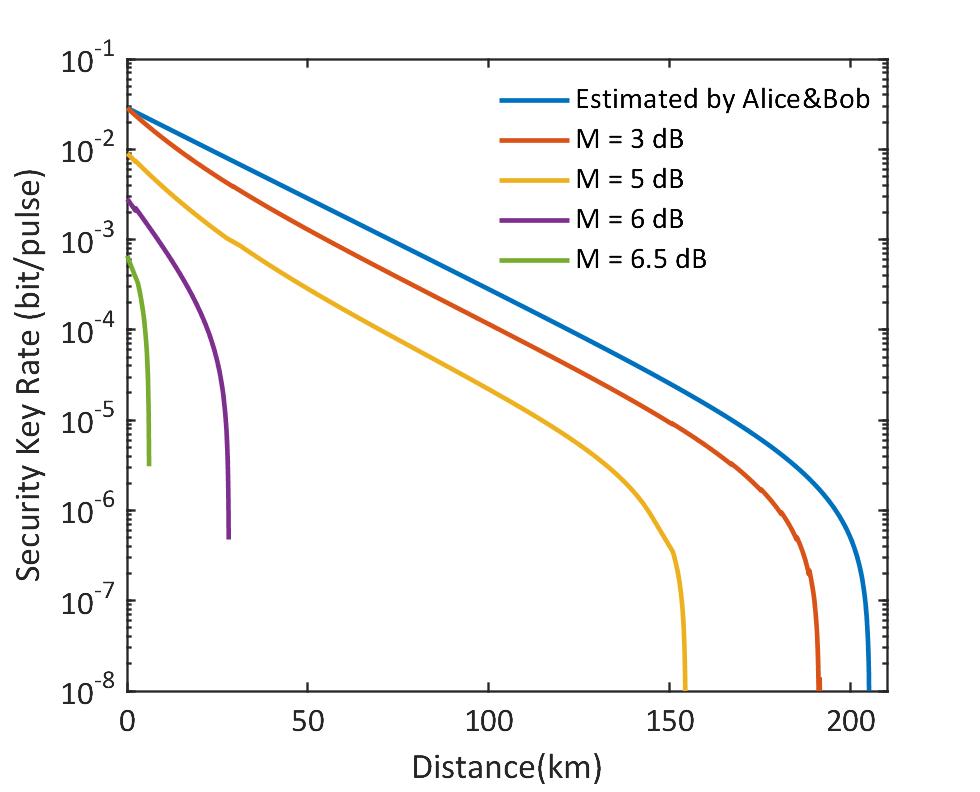}
	\caption{
		The secret key rate under different $M$ (dB). The blue line is the key rate estimated by Alice and Bob based on their measurement results. }
	\label{fig: SecurityKey}
\end{figure}
As shown in Fig. \ref{fig: SecurityKey}, the key measured by users is estimated based on the actual response rate of detectors, but the actual secure key is much smaller because of the underestimated $\Delta$. When $M$ reaches $6.5$ dB, there is no security key anymore, which means that the security can be totally broken by even the weakest irradiation power in our experiment, that is, the irradiated power of 3 nanowatts in Fig. \ref{fig:AllExpResult}a, corresponding to $M\approx 8.3$ dB.
Such a weak injection light not only reduces the requirements of C2 but also makes it more difficult to detect the existence of IPA.
  
\section{Discussion}
To counter IPA, filters such as dense wavelength division multiplexing (DWDM) modules may be insufficient because some channels don't have enough attenuation of irradiation beam (see Appendix \ref{App:IrrLoss}). Additionally, the optical power meters without a small enough power range 
are insufficient to detect the irradiation power as small as 3 nW. Fortunately, we find the commonly used isolator and circulator have a huge attenuation of irradiation beams (see Appendix \ref{App:IrrLoss}) thus they are necessary to isolate the injected beam from channels. However, the security of isolators and circulators also needs further study because they may be affected by the laser-damage attack \cite{huang2020laser, 2022ponosovaPQ}. 
Since the PE is fabrication-dependent \cite{fujiwara1993comparison}, defect engineering may be useful to reduce the PE by doping with various impurities \cite{kong2020recent}. 

In fiber systems, although we can substitute the LN-based VOA with other types \cite{huang2020laser}, the remaining high-speed modulators made of LN are unavoidable \cite{tfqkd,2022wangNP}. Therefore, state monitoring of LN devices is necessary and bias control or stabilization techniques \cite{park2022experimental,iskander2019stabilization} for MZ modulators also help to counteract the effects of IPA.
Interestingly, benefit from our research on IPA, it is a feasible countermeasure that users can also process the LN device to a controlled state by the pre-treatment technique when they get them and using it at the state that is unfavourable to IPA (such as $V_{\mathrm{B}}$ in Fig. \ref{fig:AllExpResult}b).
For the LNOI platform, the tighter optical confinement and higher crystal defect concentration contribute to stronger PE \cite{2021zhuAOPA}.  
Besides, recent researches suggest some different properties of PE in LNOI \cite{jiang2017fast,2021zhuAOPA}, such as the faster response speed and minor optical mode distortion. Thus, more comprehensive considerations on fabrication processes and structural design are necessary to mitigate PE \cite{kong2020recent,2021xuOE} in future LNOI-based integrated QKD research. 

In conclusion, we reveal PE is a new mechanism that can cause loopholes in practical QKD systems. To confirm its threat to QKD, we propose two effective techniques to enable controllable attack and proceed to execute powerful hacking on a commercial VOA. Our experimental and security analysis results show that even the weakest irradiation power (3 nW) is sufficient to cause a totally broken security. 
Considering that the PE can be induced over a wide range of wavelengths (from ultraviolet to even 1549 nm \cite{kostritskii2009photorefractive}), more types of loopholes derived from the PE may be caused and further investigation is needed. Given the importance of LN devices in QKD systems, PE-related security issues deserve more attention. Our work contributes to the development of practical security in QKD systems and opens a window for designing more secure integrated QKD systems in the future.

\begin{acknowledgments}
We would like to acknowledge Prof. Changling Zou for his valuable discussions.  This work was partially carried out at the USTC Center for Micro and Nanoscale Research and Fabrication.
This work was supported by the National Key Research and Development Program of China (2018YFA0306400); National Natural Science Foundation of China (Grants No. 61622506, No. 61575183, No. 61627820, No. 61475148, No. 62105318, No. 61822115, and No. 61675189); China Postdoctoral Science Foundation (2021M693098); Anhui Initiative in Quantum Information Technologies.
\end{acknowledgments}

\appendix

\section{More details about photorefractive effect}\label{App:PEdetail}

In lithium niobate ($\rm LiNbO_{3}$, LN) waveguide, under the illumination of an irradiation power $I_{\mathrm{ir}}$, the space-charge field caused by the inhomogeneous charge distribution is
\cite{becker1985photorefractive,howerton1992photorefractive,fujiwara1993comparison}
\begin{equation}
\begin{split}
E_{\mathrm{s}}(I_{\mathrm{ir}},t)=&[\frac{\kappa \alpha }{\sigma _{\mathrm{d}}+ \sigma _{\mathrm{ph}}}I_{\mathrm{ir}}-E_{\mathrm{s}}(I_{\mathrm{ir}},0)](1-e^{-t/\tau})\\
&+E_{\mathrm{s}}(I_{\mathrm{ir}},0),
\end{split}
\label{eq:Est}
\end{equation}
where $\alpha$ is the absorption coefficient of the donor or acceptor centers,  $\sigma _{\mathrm{d}}$ and $\sigma _{\mathrm{ph}}$ are the dark conductivity and photoconductivity, respectively. The $\kappa$ is the Glass constant, which is related to the size of the waveguide. The value of $I_{\mathrm{ir}}$ here is the average spatial intensity of the irradiation beam passing through the waveguide cross-section. $\tau$ is the build-up time constant of the space-charge field. 
This space-charge field influences the signal beam by modulating the index of the waveguide through the linear electro-optic effect
\begin{equation}
\begin{split}
\Delta n(I_{\mathrm{ir}},t)=&\frac{n^{3}r_{33}\gamma }{2}E_{\mathrm{s}}(I_{\mathrm{ir}},t )\\=& \Delta n_{\mathrm{s}}(I_{\mathrm{ir}})(1-e^{-t/\tau }),
\end{split}
\end{equation}
where $n$ is the waveguide refractive index, $r_{33}$ is the linear electro-optic coefficient, and $\gamma $ is the overlap integral of the optical mode profile and the space-charge field. This is the time response characteristic of the photorefractive effect (PE), that is, the photorefractive index change saturates with increasing irradiation time due to the time response function of type $1-e^{-t/\tau}$.

Due to the outstanding electro-optical modulation characteristics, LN waveguide usually works in the presence of the external electric field, which interacts with the photogenerated charges and also affects $E_{\mathrm{s}}$. 
Since the build-up time constant $\tau$ is on the order of $10^{3}$ seconds for both Ti-indiffusion and proton-exchanged LN waveguide \cite{fujiwara1992photorefractive}, the influence of high-speed radio frequency signals on photogenerated carriers is averaged out and only leaves its direct current component, which can be written as an external static electric field $E_{\mathrm{app}}$. Therefore, The saturated photorefractive index change $\Delta n_{\mathrm{s}}(I_{\mathrm{ir}})$ should be 
\begin{equation}
\begin{split}
\Delta n_{\mathrm{s}}(I_{\mathrm{ir}})=&\frac{n^{3}r_{33}\gamma }{2}E_{\mathrm{s}}(I_{\mathrm{ir}},t ) |_{t\rightarrow \infty }\\
=&\frac{n^{3}r_{33} \gamma}{2 }(-\frac{\sigma _{\mathrm{ph}}}{\sigma}E_{\mathrm{app}}+\frac{\kappa \alpha}{\sigma} I_{\mathrm{ir}}),
\end{split}
\label{eq:thetaPEtinf}
\end{equation}
where $\sigma =\sigma _{\mathrm{d}}+\sigma _{\mathrm{ph}} $.

When the irradiation intensity is low, for instance, lower than 100 $\rm W/cm^{2}$ (or about 7 $\upmu$W) for LN waveguide made by Ti-indiffusion in ref \cite{fujiwara1993comparison}, which assumes that the photoconductivity varies linearly with the irradiation intensity
\begin{equation}
\sigma _{\mathrm{ph}}=a\alpha I_{\mathrm{ir}},
\label{eq:sigmaphlower}
\end{equation}
where $a=\frac{e}{h\nu }\mu \tau _{0}\phi $ is a constant, which is related to the the electronic charge $e$, the photon energy $h\nu $, the electron mobility $\mu$, the quantum efficiency $\phi $ and the carrier lifetime $\tau _{0}$. Then we can obtain the saturated phase change of LN waveguide as
\begin{equation}
\begin{split}
\theta_{\mathrm{PE}}(I_{\mathrm{ir}} )=&\frac{2 \pi}{\lambda _{0}}(-\frac{a}{\kappa }L_{\mathrm{E}}\cdot E_{\mathrm{app}}+L)\cdot f(I_{\mathrm{ir}})
\\=&(-C\cdot E_{\mathrm{app}}+D)\cdot f(I_{\mathrm{ir}}).
\end{split}
\label{eq:thetaPEinf}
\end{equation}
Here we introduce $C = 2 a \pi L_{\mathrm{E}}/\kappa \lambda _{0}$ and $D = 2 \pi L/\lambda _{0}$. $\lambda_{0}$ is the free-space wavelength, $L$ is the arm length of MZI , $L_{\mathrm{E}}$ is the length of the waveguide modulated by the electric field. In this equation, $ f(I_{\mathrm{ir}})$ is the saturated photoinduced refractive index changed only related to the irradiation intensity \cite{fujiwara1993comparison}
\begin{equation}
\begin{split}
f(I_{\mathrm{ir}})=\frac{AI_{\mathrm{ir}}}{1+BI_{\mathrm{ir}}},
\end{split}
\label{eq:fmin}
\end{equation}
where $A = n^{3}r_{33} \gamma\kappa a/2\sigma _{\mathrm{d}}$ and $B = a\alpha/\sigma _{\mathrm{d}}$. We can see that the change of the photoinduced phase is related to both the applied electric field and the irradiation beam. 

At the high irradiation intensity (as before, lager than 100 $\rm W/cm^{2}$ for Ti-indiffusion waveguide), the photoconductivity can be conveniently assumed to be proportional to $(I_{\mathrm{ir}})^{1/m}$, where m is an integer greater than 1 \cite{robertson1997photorefractive}. Due to the very large $I_{\mathrm{ir}}$, we have $\sigma _{\mathrm{ph}}\gg \sigma _{\mathrm{d}}$, and
\begin{equation}
f(I_{\mathrm{ir}})=\frac{A}{B}(I_{\mathrm{ir}})^{1-1/m}.
\label{eq:fmax}
\end{equation}
Then Eq. \ref{eq:thetaPEinf} should be rewritten as 
\begin{equation}
\begin{split}
\theta_{\mathrm{PE}}(I_{\mathrm{ir}})=-\frac{A}{B}\cdot C\cdot E_{\mathrm{app}}+D\cdot f(I_{\mathrm{ir}}).
\end{split}
\label{eq:thetaPEinfbig}
\end{equation}

From both Eq. \ref{eq:fmin} and Eq. \ref{eq:fmax}, we can see that the saturated photoinduced refractive index changed also tend to saturate with the increase of the irradiation power. Therefore, the PE does not become infinitely large with the increase of $I_{\mathrm{ir}}$. This is the saturation characteristic of PE.

In practical application, researchers can dope different impurities to enhance or weaken the PE in LN crystal to make devices suitable for different application scenarios \cite{bryan1984increased,stepic2006beam,ruter2010observation}. However, the doping techniques are not mature enough and increase costs, thus have not been widely used in commercial devices. 

In addition to doping, the photorefractive sensitivity depends on the specific fabrication procedure, including Ti-indiffusion technology, proton-exchanged, annealed proton-exchanged (APE), reverse proton-exchanged (RPE), and soft proton exchange (SPE) \cite{bazzan2015optical}.
In Eq. \ref{eq:thetaPEinf} and Eq. \ref{eq:fmin}, the contribution of dark conductivity to the PE is dominant at a lower irradiation power. So the photorefractive sensitivity of devices fabricated by Ti-indiffusion technology is higher than that of other processes because of lower dark conductivity. Benefitting from the increased dark conductivity induced by the proton exchange process, proton-exchanged waveguides exhibit the highest robustness to PE but are penalized by degradation of electro-optic and nonlinear optic properties. APE, RPE, and SPE can partially recover the degradation of material properties during fabrication but are accompanied by an increase in photorefractive sensitivity \cite{mondain2020photorefractive,bazzan2015optical,fujiwara1993comparison}. 
At high irradiation power, the influence of PE can be characterised by Eq. \ref{eq:fmax} and Eq. \ref{eq:thetaPEinfbig}.
This means the saturated index change is independent of dark conductivity, so the photorefractive index change in LN waveguide at higher irradiation power is almost the same for all these fabrication procedures \cite{fujiwara1989evolution,fujiwara1992photorefractive}.  

\section{Design of the attack module}\label{App:AttModule}
\begin{figure}[htbp]
	\centering\includegraphics[width=\linewidth]{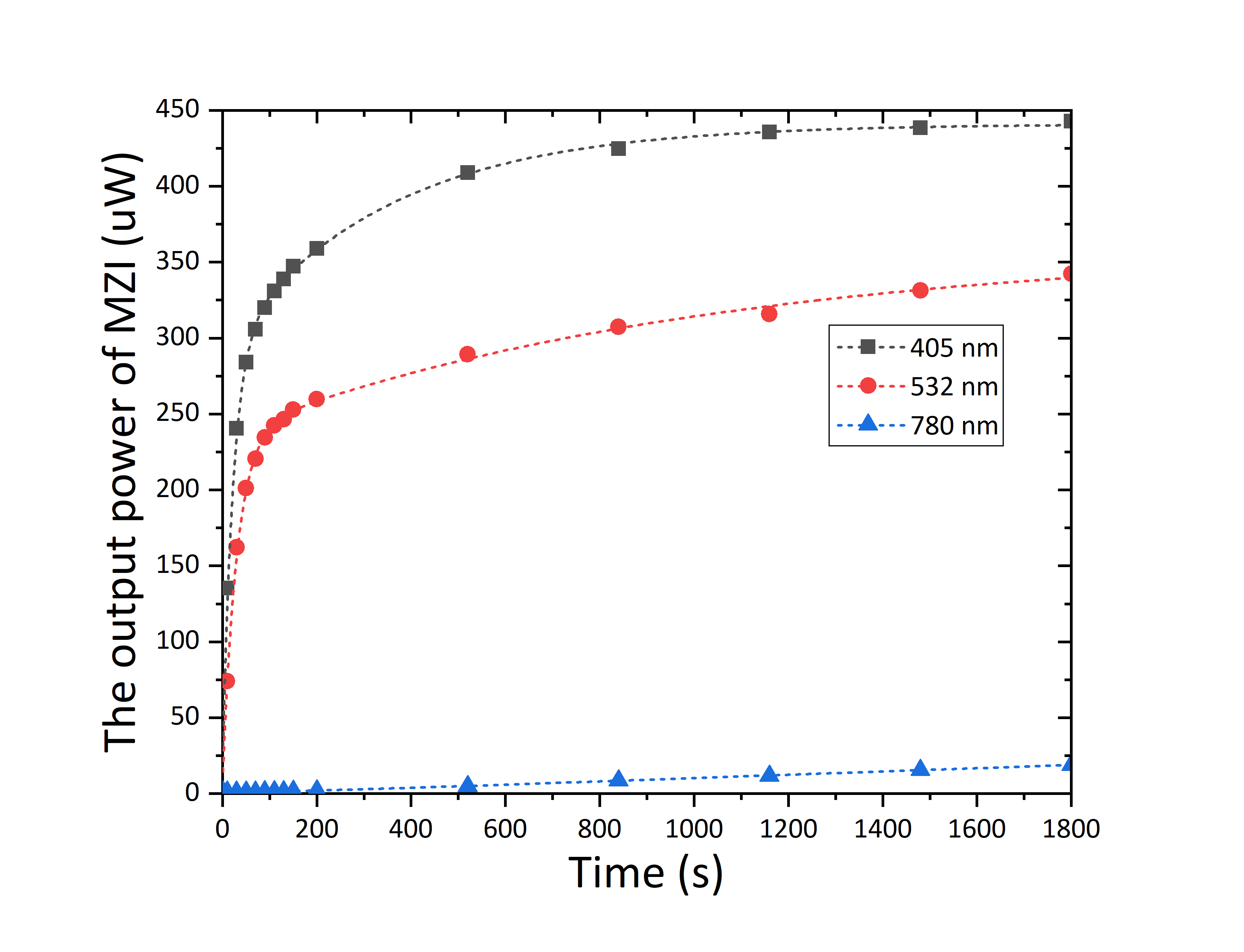}
	\caption{The wavelength dependence of PE. The time response curve of VOA under irradiation beams at 405 nm (black), 532 nm (red), and 780 nm (blue), respectively. The Y-axis is the output power of VOA and the X-axis is the injection duration. The VOA is initialized to the maximum attenuation before each test.}
	\label{fig:Dwavelength}
\end{figure}

The selection of irradiation wavelength depends on the photorefractive sensitivity of LN material. Here we test the wavelength dependence of PE. To compare the effect of PE, the irradiation power is set to 16 $\upmu$W for each wavelength and the results are shown in Fig. \ref{fig:Dwavelength}. The irradiation beam at 405 nm has the fastest response and most substantial effect. This phenomenon is consistent with the result of Fujiwara \emph{et al.} \cite{fujiwara1989wavelength,fujiwara1992photorefractive} due to LN has higher photorefractive sensitivity at a shorter wavelength. Additionally, the PE-based attack works in a wide range of wavelengths (from ultraviolet to even 1549 nm \cite{jiang2017fast,kostritskii2009photorefractive}, which is close to the commonly used communication wavelength). This increases the difficulty of defense because a band-pass filter with a sufficient wide working range of wavelength is needed to filter the irradiation beam. Therefore, our experiment used an irradiation laser at 405 nm for a more efficient attack.

To couple the irradiation beam into the communication channel, the coupling efficiency of different commercially available fiber devices working at 1550 nm has been tested, including circulator, 50:50 BS, 90:10 BS, and 99:1 BS. We find that the circulator isolates almost all irradiation power at 405 nm. 
We measure these fiber BSs' BSR and transmission loss (TL) when used in the attack module at 405 nm and 1550 nm. Table \ref{tab:BS} shows the measure result. The TL includes both insertion loss and splitting loss. For irradiation beam at 405 nm, in addition to the insertion loss, about 90 percent of light passes through the bar port for all BSs. For the two unbalanced BSs, the bar port always has a larger splitting ratio. Therefore, we select the bar port to transmit the signal light to reduce the TL at 1550 nm. As a result, the TL at 405 nm for all unbalanced BS is measured when the irradiation beam is coupling through the cross port. For 50:50 BS, the TL of the two output ports is the same at 1550 nm so that we can transmit the attack beam through the bar port, and this is why the TL of the 50:50 BS is much lower than the other two BSs.
\begin{table}[htbp]
	\centering
	\caption{When used in the attack module, the beam splitting ratio (BSR) and the transmission loss (TL) of each BS at 405 nm and 1550 nm.}
	\begin{tabular}{cccc}
		\hline
		\hline
		Types of BS & \multicolumn{1}{c}{\begin{tabular}[c]{@{}c@{}}TL \\@1550 nm (dB)\end{tabular}} & \multicolumn{1}{c}{\begin{tabular}[c]{@{}c@{}}BSR \\@405 nm\end{tabular}} & \multicolumn{1}{c}{\begin{tabular}[c]{@{}c@{}}TL \\@405 nm (dB)\end{tabular}}\\
		\hline
		50:50 1*2 BS & 3.4 & 88:12 & 7.41 \\
		90:10 1*2 BS & 0.79 & 91:9 & 13.49 \\
		99:1 1*2 BS & 0.41 & 92:8 & 15.5 \\
		\hline
		\hline
	\end{tabular}
	\label{tab:BS}
\end{table}

\begin{figure}[htbp]
	\centering\includegraphics[width=0.85\linewidth]{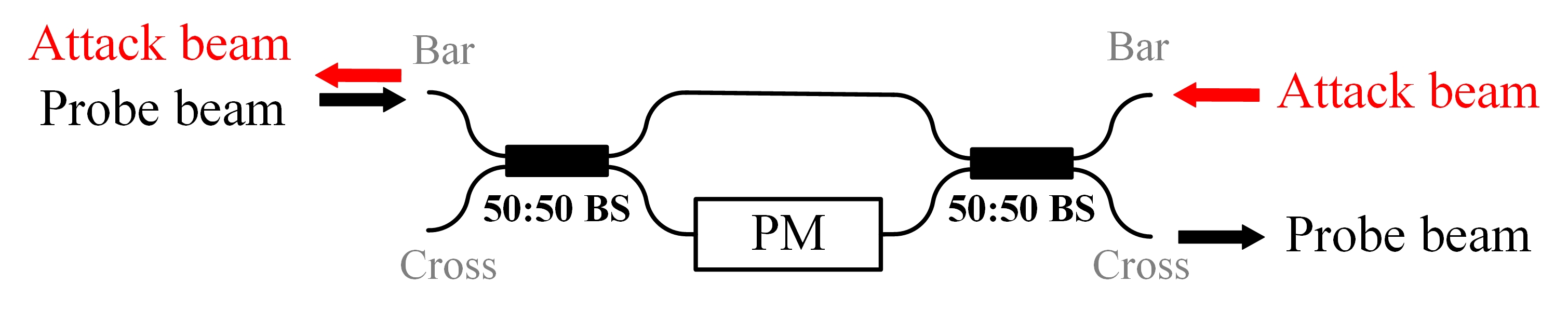}
	\caption{The MZI-based scheme includes two 50:50 BSs and a phase modulator (PM). Almost all attack power goes through the bar path of each BS. The PM is used to interfere all the signal power into the cross port of MZI.}
	\label{fig:AttackModule}
\end{figure}

In an actual communication system, the channel loss caused by the eavesdropper should be as small as possible. The best strategy to inject the attack beam is using a Mach-Zehnder interferometer (MZI) based on two 50:50 BSs. Both bar ports of BSs are used to transmit the irradiation beam, and both the cross ports are used to pass the signal light. The schematic diagram of this method is shown in Fig.\ref{fig:AttackModule}. The phase difference in MZI is modulated to $\pi$ by the phase modulator to make the signal beam output through the cross port. The TL of MZI at 405 nm we measure is about -8.31 dB, which is still lower than the other two unbalanced BSs. Theoretically, the TL at 1550 nm of this scheme is negligible. In our experiment, for the sake of simplicity, the attack beam is reversely fed into the MZI by a 50:50 BS. It is worth noting that this does not affect the main conclusions of this article. 

\section{Experiment method}\label{App:Expmethod}
\begin{figure}[htbp]
	\centering\includegraphics[width=0.85\linewidth]{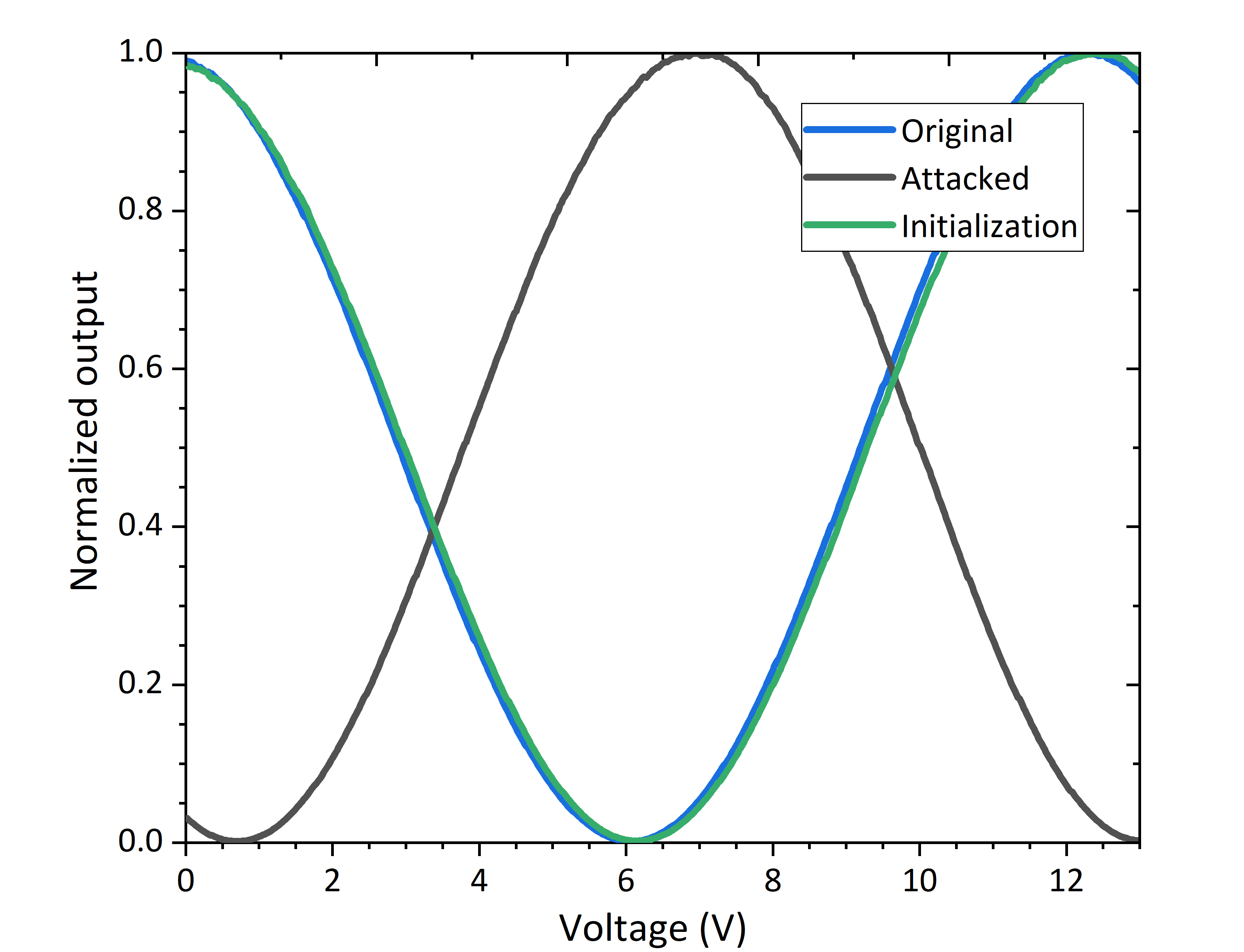}
	\caption
	{
		The initialization of VOA is shown here. VOA's original and attacked voltage curves are shown as the blue and black lines, respectively. Using the initialization method, the voltage curve can be initialized to the green line, which is almost the same as the original one.
	}
	\label{fig:VoltageCurve2}
\end{figure}

To ensure the consistency of our experiments, the comparison of PE between different powers or wavelengths should be performed in the same state of VOA. However, as shown in Fig. \ref{fig:VoltageCurve2}, the static bias of VOA is changed after each injection. Hence it is essential to recover the state of VOA to the same initial state before each experiment. The recovery of the carrier distribution depends on the dark conductivity \cite{glass1980absorption} of LN and usually persists for a long time \cite{1978glassOE}. Treating samples by heating, annealing, or uniform illuminating with a high-intensity halogen bulb can accelerate this process \cite{moretti2005temperature,mondain2020photorefractive}. Essentially, these methods erase the space charge distribution (restoring the LN material to the state before injection) by enhancing the charge mobility. But here, since the commercial device is already packaged, we cannot use these initialization methods. 

Fortunately, although the space charge distribution cannot be erased, it can be redistributed to the same state before each experiment based on our pre-treatment technique. Here, we initialize our VOA by injecting the irradiation beam at 4.39 $\upmu$W until saturation under a 0 V external electric field. As we expected, the result of initializing is the green line in Fig. \ref{fig:VoltageCurve2}, which is almost the same as the original one. All of our tests are initialized by this method.

\section{Loss of irradiation beam in devices working at 1550 nm}\label{App:IrrLoss}
To explore countermeasures against IPA, we measure the loss of irradiation beam in some commonly used devices in QKD systems. In Tab. \ref{tab:Loss}, the transmittance losses (TL) of these three irradiation wavelengths in single-mode fiber in compliance with the ITU-T G.652.D and the insertion loss of isolator, circulator, and some channels of dense wavelength division multiplexing (DWDM) module are shown. Considering the nanowatt magnitude irradiation power, we think their TL is acceptable because Eve can attack at one kilometer (which is an available distance) from the users. Moreover, QKD systems always output with narrowband optical filters such as DWDM aligned on the ITU grid. But we find the loss of the irradiation beam at 405 and 780 nm is also acceptable. For example, the total loss is 46 dB at 405 nm when a QKD system output at 1550.92 nm and Eve attacks at one kilometer. It requires 120 $\upmu$W irradiation power and this is easily available.
\begin{table}
	\centering
	\caption{The transmittance loss (TL) of irradiation beams in single-mode fiber and the insertion loss of isolator, circulator, and dense wavelength division multiplexing (DWDM) module working at 1550 nm. For DWDM aligned on the ITU grid, channel 33 (C33) and 35 (C35) are aligned to 1550.92 nm and 1549.32 nm, respectively.}
	\resizebox{\linewidth}{!}{ 
		\begin{tabular}{cccccc}
			\hline
			\hline
			\multirow{2}{*}{\begin{tabular}[c]{@{}c@{}}Wavelength \\(nm)\end{tabular}} & \multirow{2}{*}{\begin{tabular}[c]{@{}c@{}}TL\\(dB/km)\end{tabular}} & \multicolumn{2}{c}{DWDM} & \multirow{2}{*}{\begin{tabular}[c]{@{}c@{}}Isolator\\(dB)\end{tabular}} & \multirow{2}{*}{\begin{tabular}[c]{@{}c@{}}Circulator\\(dB)\end{tabular}}  \\
			&    & C33 (dB) & C35 (dB) &   &  \\
			\hline
			405 & 13 & 33 & 61 & $>78$ & $>78$ \\
			532 & 14 & 71 & $>78$ & $>78$ & $>78$ \\
			780 & 3  & 15 & 31  &  58   &   73   \\
			\hline
			\hline              
		\end{tabular}
	}
	\label{tab:Loss}
\end{table}

Learning from the Trojan-horse attack \cite{2021tanPRA, PhysRevX.5.031030}, the commonly used defenders such as the isolator and circulator are necessary to isolate the injected beam. Fortunately, they have very large losses of the irradiation wavelengths we used. Take the 405 nm irradiation beam as an example, the total loss is large than 90 dB when Eve attacks at one kilometer from the users regardless of whether the isolator or circulator is used. This means that even considering the weakest case (3 nW in our experiment), the power required by Eve already reaches the watt magnitude. To defend the Trojan-horse attack, the required isolation in different QKD systems is usually greater than 100 dB \cite{2021tanPRA, PhysRevX.5.031030}, which corresponds to at least three of the isolators we tested here (36 dB for each at 1550 nm). This contributes to a huge attenuation and makes the injection of irradiation beams almost impossible. However, the security of isolators and circulators can also be affected by the laser-damage attack thus further research is still necessary \cite{huang2020laser, 2022ponosovaPQ}.

\section{Impact of other dimensions on IPA}
In addition to the irradiation power, the applied electric field and the injection duration, we further investigate the effect of polarization and wavelength of irradiation beam on IPA.
\begin{figure}[htbp]
	\centering\includegraphics[width=\linewidth]{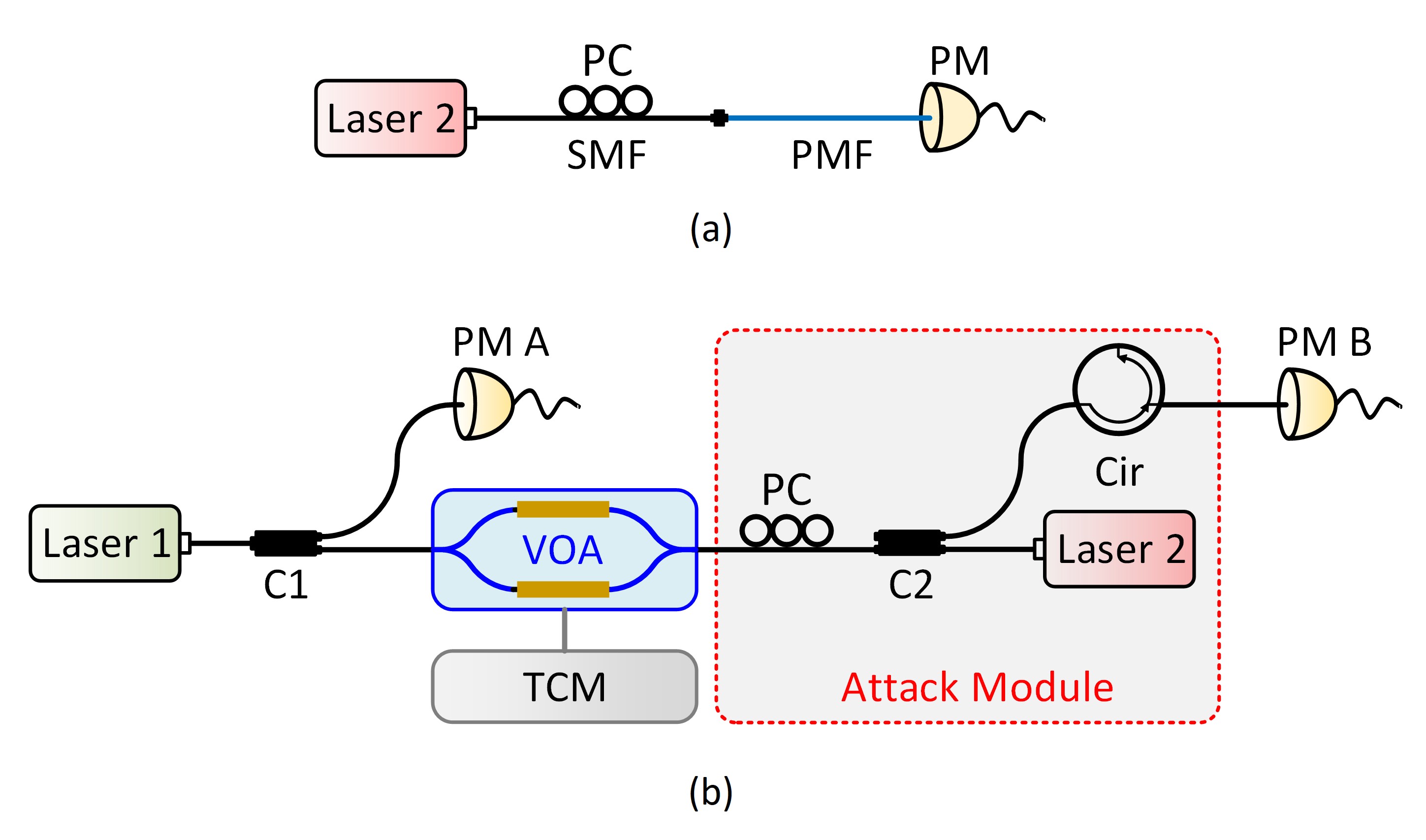}
	\caption{Sketch of the experimental setups. PC: polarization controller. SMF: single-mode fiber, which is the black line. PMF: polarization-maintaining fiber, which is the blue line. The remaining devices are the same as Fig. 2 in the main text. (a) Polarization-dependent loss test of irradiation beam passing through the polarization maintaining fiber. (b) Polarization dependence test of irradiation beam on PE.}
	\label{fig:SetupR1}
\end{figure}

The mechanism of how the irradiation polarization affects the PE is complex \cite{2006gunter} and may include various factors such as the change of coupling efficiency of the irradiation beam and the absorption process of irradiation photons. Therefore, further in-depth analysis is needed in our future research. Here, to learn how much the effect of polarization on M is, we investigate it experimentally. 
		
\begin{figure}[htbp]
	\centering\includegraphics[width=0.85\linewidth]{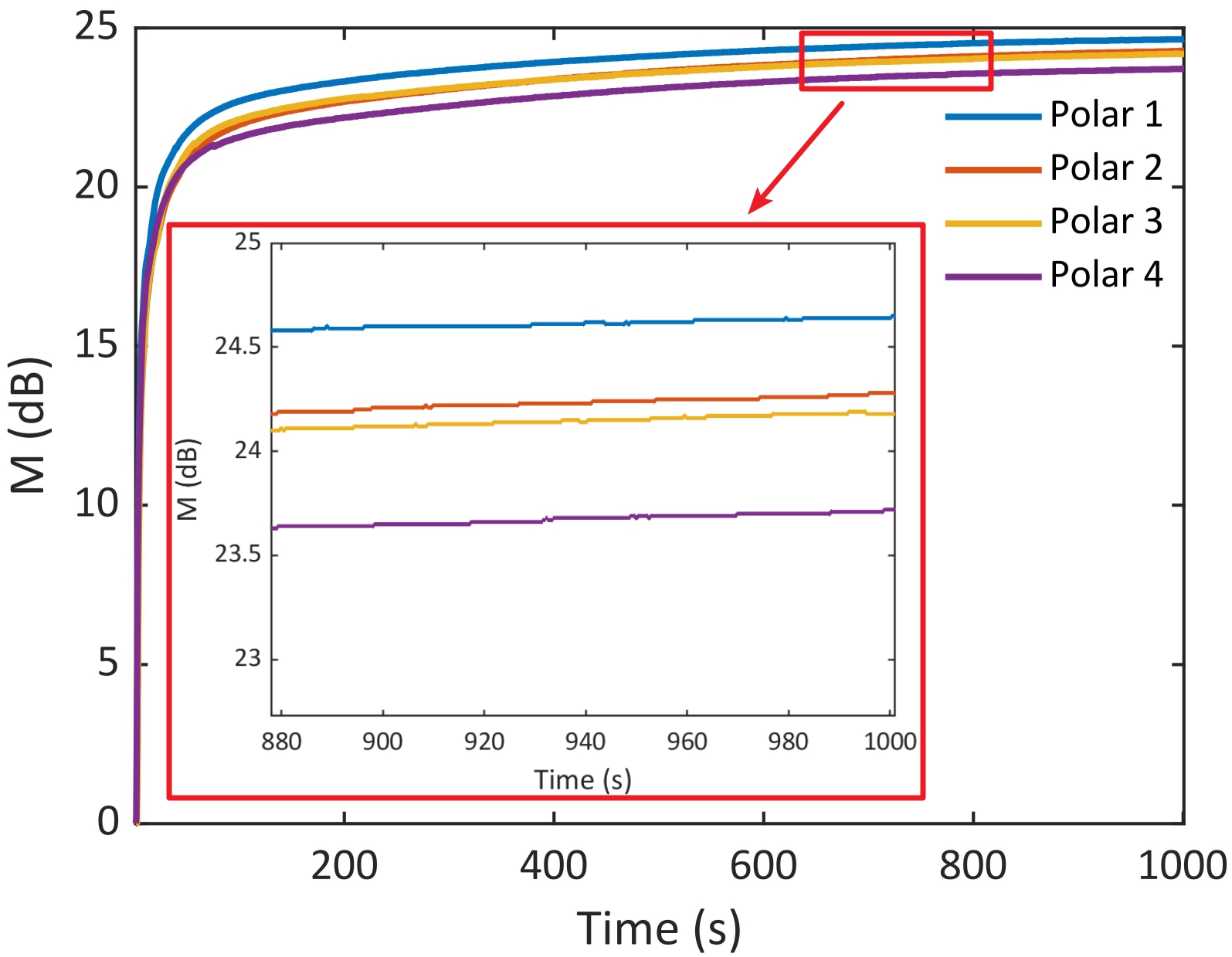}
	\caption{The change of VOA’s output at different irradiation polarization. The blue (Polar 1) and purple (Polar 4) line corresponds to the polarization that the irradiation beam has the maximum and minimum transmittance through the PMF, respectively. The red (Polar 2) and yellow (Polar 3) are the other two polarization states that are different from the blue and purple ones.}
	\label{fig:Dpolar_v2}
\end{figure}

We first test the polarization-dependent loss (PDL) of the irradiation beam passing through a polarization-maintaining fiber (PMF). Since the output fiber of the VOA is a PMF, we should know the PDL of the irradiation beam in PMF to deduce the change in the irradiation power into the VOA. As shown in Fig. \ref{fig:SetupR1}(a), the irradiation beam’s polarization is modulated by a polarization controller (PC) and then fed into a PMF. The transmittance is recorded by the optical power meter (PM) while randomly adjusting the PC. According to our test, the PDL is just about 0.1 dB, that is, the polarization of irradiation beam doesn’t affect the irradiation power fed into the VOA. Then we test the effect of the irradiation beam at different polarization by using the PC to control its polarization. The experiment setups are shown in Fig. \ref{fig:SetupR1}(b) and the results are shown in Fig. \ref{fig:Dpolar_v2}. We modulate four different polarization states, the “Polar 1” and “Polar 4” corresponds to the polarization state that has the maximum (about -18.78 dBm) and minimum (about -18.92 dBm) irradiation power cross the PMF, respectively. The “Polar 2” and “Polar 3” are the other two different polarization states. As the results show here, there is only a 0.93 dB effect on M for different polarization states and it is almost negligible for IPA.

\begin{figure}[htbp]
	\centering\includegraphics[width=0.85\linewidth]{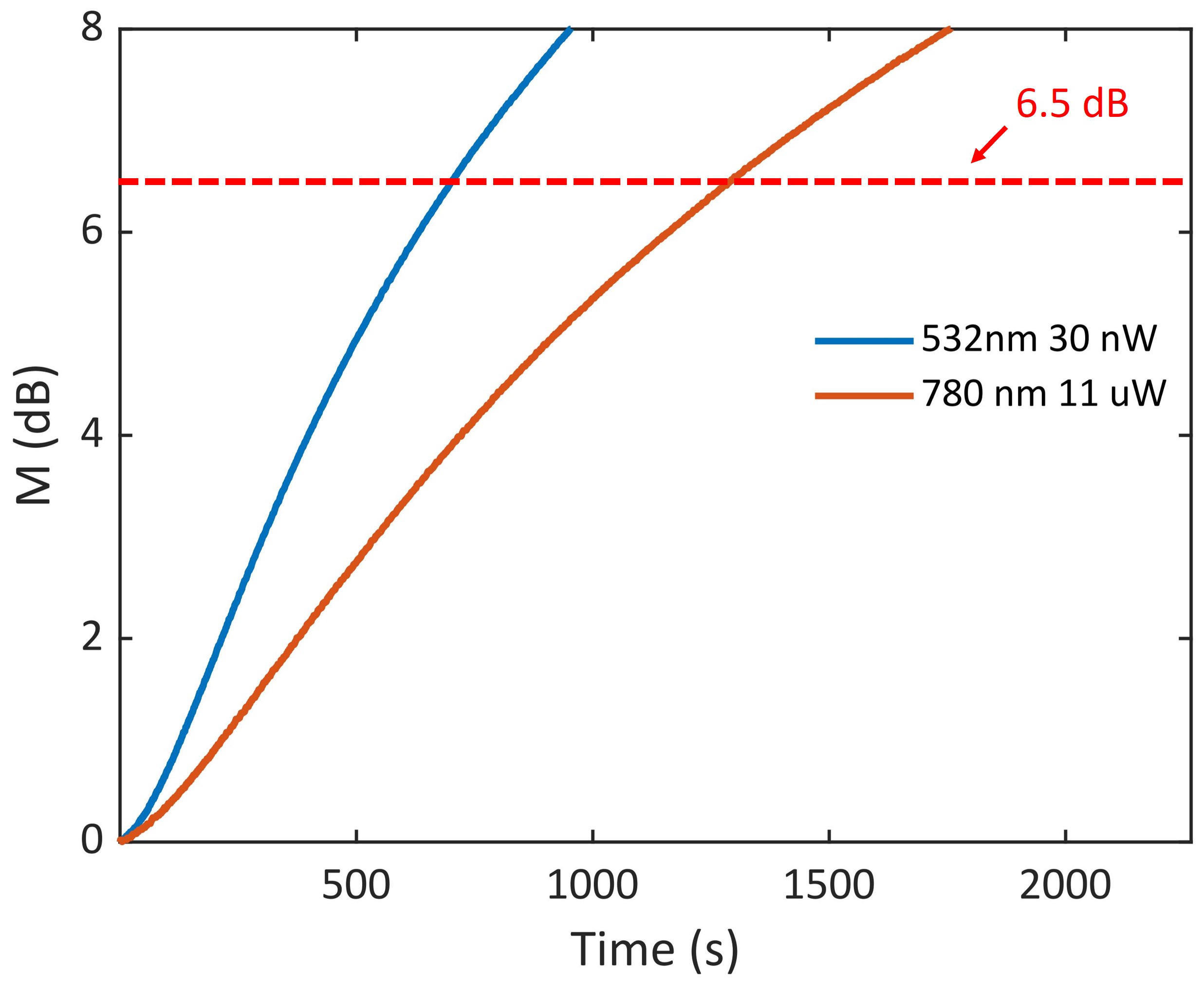}
	\caption{The effect of the irradiation beam at 532 and 780 nm.}
	\label{fig:Dwavelength3}
\end{figure}

The attack effect of different irradiation wavelengths at the same power has been given in Fig. \ref{fig:Dwavelength}, here we investigate the power required for a successful attack at each wavelength. As shown in Fig. \ref{fig:Dwavelength3}, the dotted line represents the magnification (M = 6.5 dB according to the security analysis in Fig. 4 at the main text) required to steal the security key completely and this can be achieved by both the 30 nW of 532 nm irradiation beam and the 11 $\upmu$W of 780 nm irradiation beam. However, there is no observable phenomenon for irradiation beam at 1550 nm, even at a power of up to 4 mW. Higher power is temporarily unavailable in our apparatus but we will investigate the relevant issues deeply in our subsequent work. Here, we can gain relevant knowledge from the work of Kostritskii \cite{kostritskii2009photorefractive}, that is, the required power is larger than 100 mW for wavelengths near 1.5 $\upmu$m. This value can only be used as a reference since PE is dependent on both the fabrication process and the structure of the LN devices.

\bibliography{sample}

\end{document}